\documentclass{article}
\usepackage{srcltx}
\usepackage{amsmath}
\usepackage{amssymb}
\usepackage{amsfonts}
\usepackage{latexsym}
\usepackage{textcomp}
\usepackage{appendix}
\usepackage{multirow}
\usepackage{booktabs}
\usepackage{url}
\usepackage{array}
\usepackage{marvosym}
\usepackage{graphics}
\usepackage{graphicx}
\usepackage{subfigure}
\usepackage{epsfig}
\usepackage{natbib}

\title{The impact of the financial crisis on the long-range memory of European corporate bond and stock markets}

\author{Lisana B. Martinez \\   \scriptsize{Instituto de Investigaciones Econ\'omicas y Sociales del Sur, UNS-CONICET.} \\  \scriptsize{12 de Octubre y San Juan, B8000CTX Bah\'{\i}a Blanca, Argentina.} \\ \scriptsize{Universidad Provincial del  Sudoeste (UPSO).} \\ \scriptsize{ Alvarado 328, B8000CJH Bah\'ia Blanca, Argentina} \and M. Bel\'en Guercio \\  \scriptsize{Instituto de Investigaciones Econ\'omicas y Sociales del Sur, UNS-CONICET.} \\  \scriptsize{12 de Octubre y San Juan, B8000CTX Bah\'{\i}a Blanca, Argentina.} \\ \scriptsize{Universidad Provincial del  Sudoeste (UPSO).} \\ \scriptsize{ Alvarado 328, B8000CJH Bah\'ia Blanca, Argentina} \and  Aurelio F. Bariviera\\ \scriptsize{Department of Business, Universitat Rovira i Virgili, Av. Universitat 1, 43204 Reus, Spain} \\ \scriptsize{\ttfamily aurelio.fernandez@urv.net}   \and Antonio Terce{\~{n}}o \\ \scriptsize{Department of Business, Universitat Rovira i Virgili, Av. Universitat 1, 43204 Reus, Spain}}

\begin{document}
\maketitle

\begin{abstract}
This paper investigates the presence of long memory in corporate bond and stock indices of
six European Union countries from July 1998 to February 2015. We compute the Hurst
exponent by means of the DFA method and using a sliding window in order to measure long range dependence. We detect that Hurst exponents behave differently in the stock and bond
markets, being smoother in the stock indices than in the bond indices. We verify that the level
of informational efficiency is time-varying. Moreover we find an asymmetric impact of the 2008 financial crisis in the fixed income and the stock markets, affecting the former but not the latter. Similar results are obtained using the R/S method.
 \\
\textsl{\textbf{JEL classification:} G14,C40.} \\
\textsl{\textbf{Keywords:} Hurst, DFA, corporate bond indices, stock indices, financial crisis.}
\end{abstract}

\section{Introduction}
The efficient market hypothesis (EMH) is the cornerstone of financial economics. The origins can be trace back to \cite{Gibson}, who wrote that prices of publicly traded shares ``may be regarded as the judgment of the best intelligence regarding them''. Some years later \cite{Bachel} developed the first mathematical model concerning the behavior of stock prices. Nevertheless, the study of informational efficiency begun to be studied systematically in the 1960s, when financial economics emerged as a new area within economics. The classical definition of \cite{Fama76} says that a market is informationally efficient if it ``fully reflect all available information''. Therefore, the key element in assessing efficiency is to determine the appropriate set of information that impels prices. Following \cite{Fama70}, informational efficiency can be divided into three categories: (i) weak efficiency, if prices reflect the information contained in the past series of prices, (ii) semi-strong efficiency, if prices reflect all public information and (iii) strong efficiency, if prices reflect all public and private information. As a corollary of the EMH, one cannot accept the presence of long memory in financial time series, since its existence would allow a riskless profitable trading strategy. If markets are informationally efficient, arbitrage prevent the possibility of such strategies. 

An important part of the literature focused its attention on studying the long-range dependence in stock markets. In spite of the fact that the market for corporate bonds is very important in volume and is a reference for portfolio managers, who see in these bonds an attractive way of investment, it has been overlooked in empirical studies. 

The aim of this paper is to analyze the evolution of the long memory in returns of corporate bonds indices and stock indices of six EU countries. This article contributes to the literature on EMH in four important aspects. First, we expand the empirical studies by analyzing the long memory of corporate bonds indices, since they have been less studied than other instruments of capital markets. Second, we perform a comparative analysis of the fixed income and stock markets, since both are complementary in strategic asset allocation. Third, we shed light on the asymmetric impact of the 2008 financial crisis on both markets. Fourth, the time series is long enough to reach consistent conclusions.

The paper is organized as follows. Section 2 presents a literature review on long memory of stock and bond markets. Section 3 introduces the Hurst exponent as a measure for long-range dependence. Section 4 presents the data and methodology that will be used in this paper. Section 5 exposes the empirical results. Finally, Section 6 draws the main conclusions. 

\section{Literature review}

The vast majority of the studies on informational efficiency focused their attention on the analysis of stock markets, giving other financial assets less attention. The empirical literature is not conclusive regarding the existence of long-range memory in these markets. 

There are several works that, using different methodologies and samples, find evidence of long-range dependence. In particular, \cite{GreeneFielitz77} and \cite{Mills93} use the Hurst exponent to detect the presence of long memory in the US and the UK stock markets, respectively. In \cite{FF88} positive short term autocorrelation and negative long term autocorrelation is found, after examining the returns of a diversified portfolio of the NYSE. This result reinforces the idea of an underlying mean-reverting process. Long memory is also found in the Spanish stock market \cite{BlascoSantamaria96} and the Turkish stock market \cite{Kilic04}. In the same line, \cite{BarkoulasBaumTravlos00} find evidence of long memory in the weekly returns of the Athens Stock Exchange during the period 1981-1990, and suggest that the strength of the memory could be influenced by the market size. Also long memory behavior in the Greek market was found by \cite{Panas01}. \cite{CajueiroTabak04} find that developed markets are more informationally efficient than emerging markets and that the level of efficiency is influenced by market size and trading costs. \cite{CajueiroTabak05causes} relate long-range dependence with specific financial variables of the firms under examinations. \cite{ZuninoPhyB07} find that the long-range memory in seven Latin-American markets is time varying. In this line, \cite{Bariviera11} finds evidence of a time varying long-range dependence in daily returns of Thai Stock Market during the period 1975-2010 and concludes that it is weakly influenced by the liquidity level and market size. \cite{Vodenskaetal08} show that volatility clustering in the S \& P 500 index produces memory in returns. \cite{LaSpada08} find long memory in the sign of transactions but not in the signs of returns.  \cite{UrecheRangau09} investigate the presence of long memory in volatility and trading volume of the Chinese stock market.\cite{CajueiroTabakUS} present empirical evidence of time-varying long-range dependence for US interest rates. They conclude that long memory has reduced over time. Moreover, \cite{CajueiroTabak2010} find that this long-range dependence, is affected by the monetary policy. Similarly, \cite{CajueiroTabakTSIR} find long range dependence in Brazilian interest rates and their volatility, providing important implications for monetary studies. Time-varying long range dependence in Libor interest rates is found in \cite{EPJB2015,BarivieraRSTA2015}. The authors conclude that such behavior is consistent with the Libor rate rigging scandal. 

\cite{PoterbaSummers88} analyze 18 international stock markets and find negative autocorrelation in the long run. Also, \cite{CheungLai95} use the fractional differencing test for long memory by \cite{GPH83} and find evidence of long memory in 5 out of the 18 markets under study. Using a different methodology, \cite{BarkoulasBaum96} applied spectral regression to time series of 30 firms, 7 sector indices and 2 broad stock indices at daily and monthly frequency, and find evidence of long memory only in 5 of the individual firms. \cite{Wright01} compares the memory content of the time series in developed and emerging stock markets, finding that the latter exhibits short term serial correlation in addition to long-range memory.  \cite{Henry02} concludes that there is strong evidence of long-range memory in the Korean market and some weak evidence on the German, Japanese and Taiwanese markets, after analyzing monthly returns of nine stock markets. Also, \cite{Tolvi03} uses a sample of 16 stock markets of OECD countries and finds evidence of long memory only in 3 of them and \cite{KasmanTurgutAyhan09} find that among the four main central European countries (Czech Republic, Hungary, Poland and Slovak Republic), only the last one exhibits long memory. 
 \cite{Cheong2010} compute the Hurst exponent by means of three heuristic methods and find evidence of long memory in the returns of five Malaysian equity market indices. This study finds that the Asian economic crisis affected the extent of long-range memory of the Malaysian stock market.

With respect to the fixed income market, \cite{Carbone04} find local variability of the correlation exponent in the German stock and sovereign bond markets. \cite{BaGuMa12} find empirical evidence of long memory in corporate and sovereign bond markets and detect that the current financial crisis affects more the informational efficiency of the corporate than sovereign market. \cite{Zunino2012}, using the complexity-entropy causality plane for a sample of thirty countries, find that informational efficiency is related to the degree of economic development. Recently, \cite{BaGuMa14} find that the long range memory of corporate bonds at European level are affected unevenly during the financial crisis. In particular, sectors closely related to financial activities were the first to exhibit a reduction in the informational efficiency.

There are some works that find no evidence of long memory in the financial time series. Among others we can cite \cite{Lo91}, in the returns of US stocks, and \cite{GrauCarles05} in the stock indices of US, UK, Japan and Spain.

As we can appreciate, the empirical studies on corporate bond markets are less abundant and more recent than those on stock markets. The importance of fixed income instruments in the composition of investment portfolios and in firm financing, gives a rationale for our study.

\section{Long range dependence}

One of the most common and classic measures of long-range dependence was proposed by \cite{Hurst51}. The Hurst's exponent $H$ characterizes the scaling behavior of the range of cumulative departures of a time series from its mean. There are several methods (both parametric and non parametric) to calculate the Hurst exponent. For a survey on the different methods for estimating long range dependences see \cite{Taqqu95} and \cite{Montanari99}. Among these methods there is the $R/S$ analysis, used in \cite{Hurst51} and described in depth by \cite{Mandel68} and \cite{MandelbrotWallis69}. This method uses the range of the partial sums of deviations of a time series from its mean, rescaled by its standard deviation. If we have a sequence of continuous compounded returns $\{r_1, r_2, \dots , r_{\tau} \}$, $\tau$ is the length of the estimation period and $\bar{r}_{\tau}$ is the sample mean, the the $R/S$ statistic is given by
\begin{equation}
(R/S)_\tau \equiv \frac{1}{s_\tau} \left[ \max_{1\leq t\leq\tau}\sum_{t=1}^\tau (r_t-\bar{r}_{\tau})- \min_{1\leq t\leq\tau}\sum_{t=1}^\tau(r_t-\bar{r}_{\tau})\right]
\label{eq:RS}
\end{equation}
where $s_{\tau}$ is the standard deviation
\begin{equation}
s_\tau\equiv \left[\frac{1}{\tau}\sum_{t}(r_t-\bar{r}_\tau)^2\right]^{1/2}
\label{sdtRS}
\end{equation}
Hurst \cite{Hurst51}, found that the following relation
\begin{equation}
(R/S)_\tau =(\tau/2)^H
\label{eq:H}
\end{equation}
is verified by many time series in natural phenomena. The use of the $R/S$ analysis in economic time series was pioneered by \cite{Mandel72}, and became very popular with the development of econophysics.

\cite{Serinaldi10} makes a critical review on the different estimation methods of the Hurst exponent, concluding that an inappropriate application of the estimation method could lead to incorrect conclusions about the persistence or anti-persistence of financial series. Although R/S method is probably one of the most extended methods to approximate long run memory in time series, it is not robust to departures from stationarity. Consequently, if the process under scrutiny exhibits short memory, the R/S statistic could indicate erroneously the presence of long memory. In this sense, \cite{Mosaic94} develops the method called Detrended Fluctuation Analysis (DFA) that is more appropriate when dealing with nonstationary data. As recognized by \cite{GrauCarles}, this method avoids spurious detection of long-range dependence due to nonstationary data. Due to this reason we select the DFA method in order to assess the existence of long memory in this paper.

The algorithm, described in detail in \cite{Peng95}, begins by computing the mean of the stochastic time series $y(t)$, for $t=1,\dots, M$. Then, an integrated time series $x(i)$, $i=1,\dots, M$ is obtained by subtracting mean and adding up to the $i-th$ element, $x(i)=\sum_{t=1}^{i}[y(t)-\bar{y}]$. Then $x(i)$ is divided into $M/m$ non overlapping subsamples and a polynomial fit $x_{pol}(i,m)$ is computed in order to determine the local trend of each subsample. Next the fluctuation function 
\begin{equation}
F(m)=\sqrt{\frac{1}{M}\sum_{i=1}^{M}{[x(i)-x_{pol}(i,m)]}^2}
\label{eq:DFA}
\end{equation}
is computed. This procedure is repeated for several values of $m$. The fluctuation function $F(m)$ behaves as a power-law of $m$, $F(m) \propto m^H$, where $H$ is the Hurst exponent. Consequently, the exponent is computed by regressing $\ln(F(m))$ onto $\ln(m)$. According to the literature the maximum block size to use in partitioning the data is $(length(window)/2)$, where \textit{window} is the time series window vector. Consequently, in this paper we use six points to estimate the Hurst exponent. The points for regression estimation are: $m=\{4, 8, 16, 32, 64, 128\}$. 

There are other methodologies to verify the presence of long-range memory. \cite{Rosso07} introduced the complexity-causality plane in order to discriminate between Gaussian from non-Gaussian processes. \cite{ZuninoCausality10} shows that this innovative approach could be used to rank stock markets according to their stage of development. In \cite{ZuninoPermutation11}, the application of the complexity-entropy causality plane was extended to the study of the efficiency of commodity prices. This method reveals that it is not only useful to produce a ranking of efficiency of different commodities, but it also allows to identify periods of increasing and decreasing randomness in the price dynamics. \cite{Zunino2012} uses this representation space to establish an efficiency ranking of different markets and distinguish different bond market dynamics and conclude that the classification derived from the complexity-entropy causality plane is consistent with the qualifications assigned to sovereign instruments by major rating companies.

\section{Data and methodology}
We calculate the Hurst exponent for daily returns of corporate bond and stock market indices of six European Union countries: Austria, Belgium, Finland, France, Italy and Luxembourg. All data used in this paper was retrieved from DataStream. The details of the series codes are in Table \ref{Datos}. The corporate bond indices are constituted by corporate bonds that are investment grade, with at least one year remaining to maturity and a minimum outstanding of  \EUR 300 million. The corporate sector consists of financial, industrial and utility companies. The stock indices are the most representative of each country.

\begin{table}
\caption{Data details.}
\begin{center}
\begin{tabular}{l l l l }
\hline
Country & Acronym & Bond Index & Stock Index  \\ \hline \\
Austria & OE & LHAACIE  & WBI \\
Belgium & BG &  LHABCIE  & BEL 20 \\ 
Finland & FN & LHAFCIE   &  OMXH \\ 
France & FR & LHAFRCE  &  CAC 40 \\
Italy  & IT & LHAICIE &   MIB   \\
Luxembourg & LX & LHALCIE &  LUXX   \\ 
\hline
\end{tabular}
\label{Datos}  
\end{center}
\end{table}

The period under study goes from 05/01/1999 until 12/02/2015 with a total of 4203 datapoints. The continuous compounded return $r_t$ is computed as follows:
\begin{equation}
r_{t+1}=\ln\left(\frac{P_{t+1}}{P_t}\right)*100
\label{eq:return}
\end{equation}

Tables \ref{descriptiveCorp} and \ref{descriptiveStock} show the descriptive statistics of daily returns, which reflect the excess of kurtosis and non normality of the data, which is a stylized fact in many financial time series (see \cite{Cont}). As could be appreciate bonds redemption yield have been positive during the whole period of analysis, except for France. Austria bonds present the highest returns, followed by Belgium and Italy. Moreover, Luxemburg and Belgium show the highest variances of the period, which means that their performance are very volatile.

The stock market yields are positive during all observation period for Austria, Finland, and France, but negative for Belgium, Italy and Luxembourg. Moreover, Finland presents greater variance, such as Luxemburg, than others selected countries. Similarly to bonds yields, these variables are not normally distributed.

\begin{table}[htbp]
  \centering
  \caption{Corporate bond indices. Descriptive statistics of daily returns.}
    \begin{tabular}{rrrrrrr}
    \toprule
    \multicolumn{1}{c}{} & \multicolumn{1}{c}{OE} & \multicolumn{1}{c}{BG} & \multicolumn{1}{c}{FN} & \multicolumn{1}{c}{FR} & \multicolumn{1}{c}{IT} & \multicolumn{1}{c}{LX} \\
    \midrule
    Observations & 4203  & 4203  & 4203  & 4203  & 4203  & 4203 \\
    Mean  & 0.0007 & 0.0005 & 0.0003 & -0.0001 & 0.0049 & 0.0023 \\
    Median & 0.0000 & 0.0000 & 0.0000 & 0.0000 & 0.0089 & 0.0000 \\
    Min   & -2.2263 & -6.2986 & -6.6244 & -1.4573 & -5.8948 & -5.5124 \\
    Max   & 2.1667 & 12.1781 & 6.6709 & 1.3761 & 6.3251 & 8.1018 \\
    Std. Deviation & 0.2095 & 0.3209 & 0.2871 & 0.1931 & 0.2798 & 0.3910 \\
    Skewness & -0.2001 & 9.7405 & -0.0091 & -0.4428 & 0.5400 & 0.0846 \\
    Kurtosis & 14.7352 & 551.7285 & 162.4596 & 6.8790 & 120.6734 & 88.0309 \\
    Jarque Bera & 24145 & 52797118 & 4452965 & 2772  & 2425167 & 1266204 \\
    \bottomrule
    \end{tabular}%
  \label{descriptiveCorp}%
\end{table}%

\begin{table}[htbp]
  \centering
  \caption{Stock market indices. Descriptive statistics of daily returns.}
    \begin{tabular}{rrrrrrr}
    \toprule
    \multicolumn{1}{c}{} & \multicolumn{1}{c}{OE} & \multicolumn{1}{c}{BG} & \multicolumn{1}{c}{FN} & \multicolumn{1}{c}{FR} & \multicolumn{1}{c}{IT} & \multicolumn{1}{c}{LX} \\
    \midrule
    Observations & 4203  & 4203  & 4203  & 4203  & 4203  & 4203 \\
    Mean  & 0.0161 & -0.0005 & 0.0094 & 0.0031 & -0.0139 & -0.0024 \\
    Median & 0.0000 & 0.0099 & 0.0000 & 0.0055 & 0.0096 & 0.0111 \\
    Min   & -8.6189 & -8.3193 & -17.4037 & -9.4715 & -8.5981 & -30.0534 \\
    Max   & 10.2799 & 9.3340 & 14.5631 & 10.5946 & 10.8769 & 33.2180 \\
    Std. Deviation & 1.2038 & 1.2673 & 1.8640 & 1.4722 & 1.5055 & 1.7081 \\
    Skewness & -0.3863 & 0.0354 & -0.3385 & 0.0042 & -0.0820 & 0.2241 \\
    Kurtosis & 12.9508 & 9.0692 & 10.2167 & 7.8733 & 7.4665 & 63.3921 \\
    Jarque Bera & 17445 & 6452  & 9201  & 4159  & 3498  & 638752 \\
    \bottomrule
    \end{tabular}%
  \label{descriptiveStock}%
\end{table}%

Departing from daily returns, we compute the Hurst exponent using the DFA method. Since we are interested in studying the dynamic behavior of the Hurst exponent along our period of study, and following \cite{CajuTabEM,CajuTab}, we estimate the Hurst exponent using a two year sliding window (500 datapoints). The selection this time-window is justified because it reflects changes in economic situation while still giving a reliable estimate of long range correlation. In order to check for the robustness of our findings we also used a 1024 datapoints sliding window, as it was previously used by \cite{ZuninoPhyB07} to study the dynamics of the Hurst exponent in Latin-American markets and by \cite{Bariviera11} to analyze the evolution of long memory in the Thai stock market. 

The rolling sample approach works as follows: we compute the Hurst exponent for the first 500 returns, then we move forward seven datapoints, compute the Hurst exponent, and continue this way until the end of data. Thus, each $H$ estimate is calculated from data samples of the same size. The last $H$ estimate covers the period from 06/03/2013 until 12/02/2015.

\section{Results}

We compute the Hurst exponent by means of the DFA method using a rolling sample. We obtain 529 Hurst exponents of each time series. Descriptive statistics are presented in Tables \ref{HCorp} and \ref{HStock}.

\begin{table}[htbp]
  \centering
  \caption{Descriptive statistics of Hurst estimates of bond indices using sliding windows.}
    \begin{tabular}{rrrrrrr}
    \toprule
    \multicolumn{1}{c}{} & \multicolumn{1}{c}{OE} & \multicolumn{1}{c}{BG} & \multicolumn{1}{c}{FN} & \multicolumn{1}{c}{FR} & \multicolumn{1}{c}{IT} & \multicolumn{1}{c}{LX} \\
    \midrule
    Observations & 529   & 529   & 529   & 529   & 529   & 529 \\
    Mean  & 0.5462 & 0.5446 & 0.5016 & 0.5549 & 0.5573 & 0.4652 \\
    Median & 0.5448 & 0.5437 & 0.5040 & 0.5528 & 0.5504 & 0.4671 \\
    Min   & 0.3954 & 0.2616 & 0.2864 & 0.4126 & 0.3623 & 0.1653 \\
    Max   & 0.7306 & 0.7702 & 0.6326 & 0.7404 & 0.7672 & 0.7246 \\
    Std. Deviation & 0.0659 & 0.0715 & 0.0539 & 0.0638 & 0.0906 & 0.0774 \\
    Skewness & 0.2714 & -0.2445 & -0.5095 & 0.3398 & 0.2372 & -0.0323 \\
    Kurtosis & 2.6587 & 3.8294 & 3.8137 & 2.6930 & 2.2301 & 3.8263 \\
    Jarque Bera & 9.0616 & 20.4317 & 37.4816 & 12.2572 & 18.0238 & 15.1405 \\
    \bottomrule
    \end{tabular}%
  \label{HCorp}%
\end{table}%

\begin{table}[htbp]
  \centering
  \caption{Descriptive statistics of Hurst estimates of stock indices using sliding windows.}
    \begin{tabular}{rrrrrrr}
    \toprule
    \multicolumn{1}{c}{} & \multicolumn{1}{c}{OE} & \multicolumn{1}{c}{BG} & \multicolumn{1}{c}{FN} & \multicolumn{1}{c}{FR} & \multicolumn{1}{c}{IT} & \multicolumn{1}{c}{LX} \\
    \midrule
    Observations & 529   & 529   & 529   & 529   & 529   & 529 \\
    Mean  & 0.5358 & 0.4891 & 0.5024 & 0.4552 & 0.4917 & 0.5317 \\
    Median & 0.5438 & 0.4869 & 0.5066 & 0.4578 & 0.4989 & 0.5568 \\
    Min   & 0.3105 & 0.2853 & 0.3791 & 0.3026 & 0.3234 & 0.2456 \\
    Max   & 0.6824 & 0.6173 & 0.6716 & 0.5625 & 0.6199 & 0.7383 \\
    Std. Deviation & 0.0692 & 0.0689 & 0.0436 & 0.0401 & 0.0464 & 0.1022 \\
    Skewness & -0.3504 & -0.2419 & -0.0217 & -0.5316 & -0.5388 & -0.6774 \\
    Kurtosis & 2.5193 & 2.6021 & 3.5265 & 3.5550 & 3.5099 & 2.7689 \\
    Jarque Bera & 15.9208 & 8.6495 & 6.1506 & 31.7088 & 31.3288 & 41.6362 \\
    \bottomrule
    \end{tabular}%
  \label{HStock}%
\end{table}%

\begin{figure}
\center \includegraphics[scale=.7]{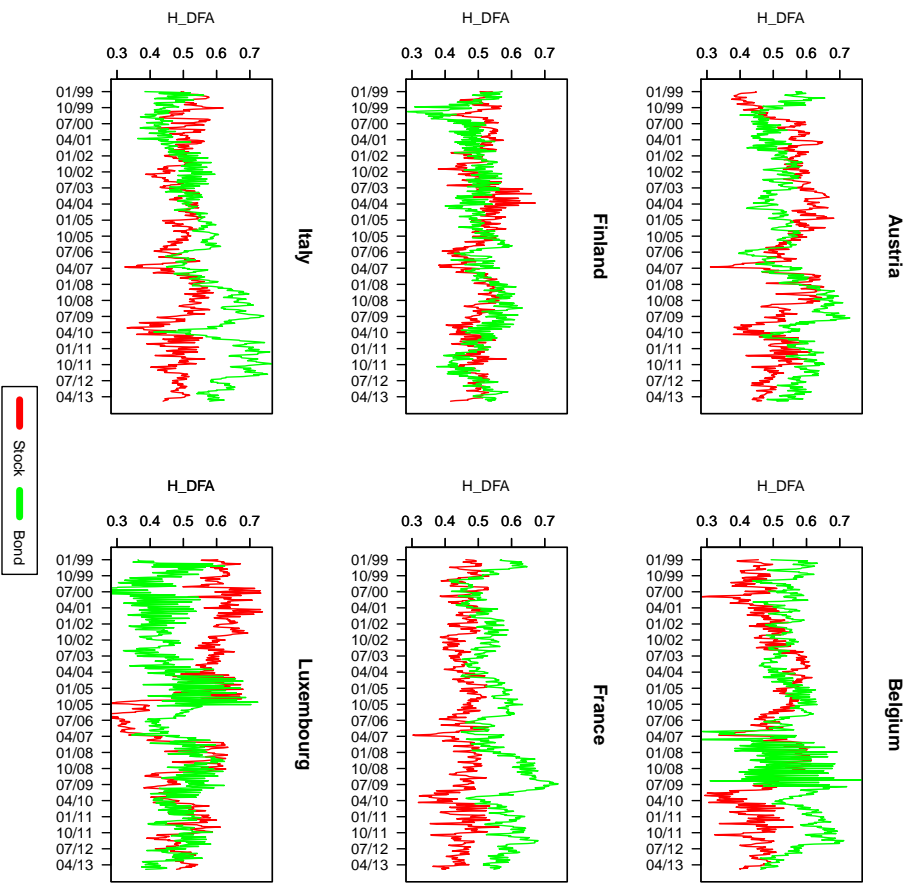}
\caption{Rolling Hurst exponent of stock indices (red) and bond indices (green). Dates in the x-axis correspond to the inital date of the respective sampling window.}
\label{Hurstsb}
\end{figure}

Figure \ref{Hurstsb} presents the evolution of the Hurst estimates of stocks and bonds for the six selected countries. As it could be observed, the lines of the former are smoother than the lines corresponding to the latter. We also observe that in almost all countries there is an increase in Hurst exponents around the window that begins in October 2005. Considering that the period of estimation of each Hurst exponent is approximately 2 years (500 datapoints), those estimates that begin in October 2005, include in their sample returns of years 2007 and after. This finding could indicate that Hurst exponents behave differently before and during the financial turmoil. A similar result is shown in \cite{Cheong2010} during the Asian crisis in the Malaysian market. 

In order to verify this hypothesis we divide the sample into two subperiods. The division is set the day Lehman Brothers filed bankruptcy, on 15/09/2008, as it is frequently used in the literature \cite{BaGuMa14,BarriosetalEU,BernothErdogan,GrammatikosVermeulen,Martinez2013}.

We carried out two analysis. First, we test the equality of the parameters of two subsamples. Second, we test if the subsamples changes in their memory endowment.

Table \ref{Mann-Whitney} shows the results of the Mann-Whitney test in order to assess whether the two subsamples have equal means. Either in bond or stock indices, the results leads to a rejection of the null hypothesis in all countries, except for corporate bonds in Finland. According to our results, the subsamples are statistically different in means, which lead us to hypothesize that the times series informational efficiency is different before and during the financial crisis.

\begin{table}[htbp]
  \centering
  \caption{Mann-Whitney test of equality of Hurst estimates before and after the crisis.}
    \begin{tabular}{rrrrrrr}
    \toprule
    \multicolumn{1}{c}{} & \multicolumn{1}{c}{OE} & \multicolumn{1}{c}{BG} & \multicolumn{1}{c}{FN} & \multicolumn{1}{c}{FR} & \multicolumn{1}{c}{IT} & \multicolumn{1}{c}{LX} \\
    \midrule
    \textbf{Bond} &       &       &       &       &       &  \\
    Mean before crisis & 0.5247 & 0.5294 & 0.4986 & 0.5342 & 0.5153 & 0.4552 \\
    Mean after crisis & 0.5919 & 0.5769 & 0.5080 & 0.5991 & 0.6468 & 0.4865 \\
    WilcoxonMannWhitney test & 11740 & 19168 & 27800 & 12991 & 5284  & 22198 \\
    p-value & \textbf{0.0000} & \textbf{0.0000} & 0.1100 & \textbf{0.0000} & \textbf{0.0000} & \textbf{0.0000} \\
    \textbf{Stock} &       &       &       &       &       &  \\
    Mean before crisis & 0.5546 & 0.5083 & 0.5098 & 0.4604 & 0.4981 & 0.5521 \\
    Mean after crisis & 0.4956 & 0.4480 & 0.4868 & 0.4441 & 0.4780 & 0.4883 \\
    WilcoxonMannWhitney test & 47306 & 45638 & 41382 & 37242 & 38175 & 46600 \\
    p-value & \textbf{0.0000} & \textbf{0.0000} & \textbf{0.0000} & \textbf{0.0000} & \textbf{0.0000} & \textbf{0.0000} \\
    \bottomrule
    \end{tabular}%
  \label{Mann-Whitney}%
\end{table}%

Table \ref{Levene} shows the results of the Levene test of Hurst estimates' homogeneity of variance. Regarding to the Hurst exponents of bond indices, we reject the null hypothesis of homogeneity of variances, for Belgium and Luxembourg. Regarding to the Hurst exponents of stock indices, the equality of variances is rejected in the time series, except in Italy.

\begin{table}[htbp]
  \centering
  \caption{Levene test of equality of Variance of Hurst exponents.}
    \begin{tabular}{rrrrrrr}
    \toprule
    \multicolumn{1}{c}{} & \multicolumn{1}{c}{OE} & \multicolumn{1}{c}{BG} & \multicolumn{1}{c}{FN} & \multicolumn{1}{c}{FR} & \multicolumn{1}{c}{IT} & \multicolumn{1}{c}{LX} \\
    \midrule
    \textbf{Bond} &       &       &       &       &       &  \\
    Std.Dev. before crisis & 0.0598 & 0.0613 & 0.0528 & 0.0548 & 0.0666 & 0.0846 \\
    Std.Dev. after crisis & 0.0539 & 0.0806 & 0.0557 & 0.0592 & 0.0669 & 0.0532 \\
    Levene test & 1.5071 & 26.4315 & 2.9708 & 1.1316 & 0.1499 & 25.0381 \\
    p-value & 0.2201 & \textbf{0.0000} & 0.0854 & 0.2879 & 0.6988 & \textbf{0.0000} \\
    \textbf{Stock} &       &       &       &       &       &  \\
    Std.Dev. before crisis & 0.0702 & 0.0660 & 0.0463 & 0.0371 & 0.0453 & 0.1135 \\
    Std.Dev. after crisis & 0.0461 & 0.0557 & 0.0322 & 0.0439 & 0.0460 & 0.0501 \\
    Levene test & 26.1964 & 12.3452 & 16.9376 & 4.1442 & 0.0035 & 68.9006 \\
    p-value & \textbf{0.0000} & \textbf{0.0005} & \textbf{0.0000} & \textbf{0.0423} & 0.9528 & \textbf{0.0000} \\
    \bottomrule
    \end{tabular}%
  \label{Levene}%
\end{table}%

Subsequently, we propose to expand the analysis, by comparing the memory content in the two subperiods. Given that we estimate Hurst exponents using sliding windows, we have a set of $H$ estimates, each of them from a sample of size 500. Consequently we can compute the confidence interval of our estimations using the t-distribution. We present the 0.999 upper and lower bounds in Table \ref{ttestbond} and Table \ref{tteststock}.

The results for the corporate bonds market are shown in Table \ref{ttestbond}. If we consider the whole period we observe that the Hurst exponents are higher than 0.5 using a 0.999 confidence interval, except for Finland and Luxembourg. These results indicate that the time series does not follow a random walk. When we split the time series into two subperiods, we find different memory dynamics. In the first subperiod, the conclusions are the same to the whole period. However, in the second subperiod we find that the mean of the Hurst exponents increased significantly. This indicates that the time series become more persistent $(H>0.5)$ during the financial crisis, even in Finland. Luxembourg is the only market that has $H<0.5$.

The results for the stock indices are shown in Table \ref{tteststock}. Looking at the whole period, the mean of the Hurst estimates are closer to 0.5, than in bond indices. Consequently, the market behaves more or less consistently with a random walk. Before the crisis, the time series were slightly persistent  $(H>0.5)$, whereas, after the crisis, the time series became anti-persistent  $(H<0.5)$. According to our results, the financial distress does not seem to affect equally the informational efficiency of the stock and the bond markets.

The memory content of stock indices, at first glance, changed from persistent to antipersistent, considering the Hurst exponent. This result could indicate a general contraction of the European stock markets, and probably, a path towards a more efficient behavior. Nontheless, with respect to the corporate bonds indices all of them have changed their informational efficiency towards an even more persistent series. This change could have been induced by the arising of a general financial turmoil in 2008. The impact of this crisis was different in the fixed income \textit{vis-\`a-vis} the stock markets. In particular, the corporate bond market was more affected by the new financial scenario, since the departure from the $H=0.5$ benchmark is larger in corporate bond market. 

A special mention is devoted to the Belgian bond market. This market is rather small compared with the rest under analysis in this paper. In particular, the Belgian financial system is bank-based, which hamper the development of alternative financial instruments, such as corporate bonds or other fixed income instruments. On the other hand, the small corporate bond market is dominated by few players, including the so-called universal banks (i.e. big banks that acts as both investment and commercial banks), which induce to a very small corporate bond outstanding, and the trading focuses on the few large banks. Hurst exponent measures long term correlations in a time series. If a market is of reduced size, or if it has very few operating agents, individual behavior could be amplified. Previous studies related informational efficiency with market size, e.g. \cite{BarkoulasBaumTravlos00}. Consequently, herd behavior, or sudden movements in the market, could contaminate the price signal with significant noise. As a result, DFA method could have problems in filtering the long range correlations. This is probably the explanation of the errant behavior in the time series of Hurst exponent in the Belgian bond market.

\begin{table}[htbp]
  \centering
  \caption{Inefficiency test of the Hurst estimates of bond indices.}
    \begin{tabular}{rrrrrrr}
    \toprule
    \multicolumn{1}{c}{} & \multicolumn{1}{c}{OE} & \multicolumn{1}{c}{BG} & \multicolumn{1}{c}{FN} & \multicolumn{1}{c}{FR} & \multicolumn{1}{c}{IT} & \multicolumn{1}{c}{LX} \\
    \midrule
    \textbf{Whole period} & \multicolumn{1}{c}{} & \multicolumn{1}{c}{} & \multicolumn{1}{c}{} & \multicolumn{1}{c}{} & \multicolumn{1}{c}{} & \multicolumn{1}{c}{} \\
    Observations & 529   & 529   & 529   & 529   & 529   & 529 \\
    Mean  & 0.5462 & 0.5446 & 0.5016 & 0.5549 & 0.5573 & 0.4652 \\
    Std. Deviation & 0.0659 & 0.0715 & 0.0539 & 0.0638 & 0.0906 & 0.0774 \\
    Std. Error & 0.0029 & 0.0032 & 0.0024 & 0.0029 & 0.0041 & 0.0035 \\
    0.999 conf. upper bound & 0.5553 & 0.5545 & 0.5091 & 0.5638 & 0.5699 & 0.4759 \\
    0.999 conf. lower bound & 0.5370 & 0.5346 & 0.4941 & 0.5461 & 0.5447 & 0.4544 \\
    \textbf{Before crisis} &       &       &       &       &       &  \\
    Observations & 360   & 360   & 360   & 360   & 360   & 360 \\
    Mean  & 0.5247 & 0.5294 & 0.4986 & 0.5342 & 0.5153 & 0.4552 \\
    Std. Deviation & 0.0598 & 0.0613 & 0.0528 & 0.0548 & 0.0666 & 0.0846 \\
    Std. Error & 0.0027 & 0.0027 & 0.0024 & 0.0024 & 0.0030 & 0.0038 \\
    0.999 conf. upper bound & 0.5330 & 0.5379 & 0.5059 & 0.5418 & 0.5245 & 0.4669 \\
    0.999 conf. lower bound & 0.5164 & 0.5209 & 0.4913 & 0.5266 & 0.5060 & 0.4434 \\
    \textbf{After crisis} &       &       &       &       &       &  \\
    Observations & 169   & 169   & 169   & 169   & 169   & 169 \\
    Mean  & 0.5919 & 0.5769 & 0.5080 & 0.5991 & 0.6468 & 0.4865 \\
    Std. Deviation & 0.0539 & 0.0806 & 0.0557 & 0.0592 & 0.0669 & 0.0532 \\
    Std. Error & 0.0024 & 0.0036 & 0.0025 & 0.0026 & 0.0030 & 0.0024 \\
    0.999 conf. upper bound & 0.5994 & 0.5881 & 0.5158 & 0.6073 & 0.6561 & 0.4939 \\
    0.999 conf. lower bound & 0.5844 & 0.5657 & 0.5003 & 0.5909 & 0.6375 & 0.4791 \\
    \bottomrule
    \end{tabular}%
  \label{ttestbond}%
\end{table}%

\begin{table}[htbp]
  \centering
  \caption{Inefficiency test of the Hurst estimates of stock indices.}
    \begin{tabular}{rrrrrrr}
    \toprule
    \multicolumn{1}{c}{} & \multicolumn{1}{c}{OE} & \multicolumn{1}{c}{BG} & \multicolumn{1}{c}{FN} & \multicolumn{1}{c}{FR} & \multicolumn{1}{c}{IT} & \multicolumn{1}{c}{LX} \\
    \midrule
    \textbf{Whole period} &       &       &       &       &       &  \\
    Observations & 529   & 529   & 529   & 529   & 529   & 529 \\
    Mean  & 0.5358 & 0.4891 & 0.5024 & 0.4552 & 0.4917 & 0.5317 \\
    Std. Deviation & 0.0692 & 0.0689 & 0.0436 & 0.0401 & 0.0464 & 0.1022 \\
    Std. Error & 0.0031 & 0.0031 & 0.0020 & 0.0018 & 0.0021 & 0.0046 \\
    0.999 conf. upper bound & 0.5454 & 0.4986 & 0.5085 & 0.4607 & 0.4981 & 0.5459 \\
    0.999 conf. lower bound & 0.5261 & 0.4795 & 0.4964 & 0.4496 & 0.4852 & 0.5175 \\
    \textbf{Before crisis} &       &       &       &       &       &  \\
    Observations & 360   & 360   & 360   & 360   & 360   & 360 \\
    Mean  & 0.5546 & 0.5083 & 0.5098 & 0.4604 & 0.4981 & 0.5521 \\
    Std. Deviation & 0.0702 & 0.0660 & 0.0463 & 0.0371 & 0.0453 & 0.1135 \\
    Std. Error & 0.0031 & 0.0030 & 0.0021 & 0.0017 & 0.0020 & 0.0051 \\
    0.999 conf. upper bound & 0.5644 & 0.5175 & 0.5162 & 0.4655 & 0.5044 & 0.5679 \\
    0.999 conf. lower bound & 0.5449 & 0.4992 & 0.5034 & 0.4552 & 0.4918 & 0.5364 \\
    \textbf{After crisis} &       &       &       &       &       &  \\
    Observations & 169   & 169   & 169   & 169   & 169   & 169 \\
    Mean  & 0.4956 & 0.4480 & 0.4868 & 0.4441 & 0.4780 & 0.4883 \\
    Std. Deviation & 0.0461 & 0.0557 & 0.0322 & 0.0439 & 0.0460 & 0.0501 \\
    Std. Error & 0.0021 & 0.0025 & 0.0014 & 0.0020 & 0.0021 & 0.0022 \\
    0.999 conf. upper bound & 0.5020 & 0.4557 & 0.4912 & 0.4502 & 0.4844 & 0.4953 \\
    0.999 conf. lower bound & 0.4892 & 0.4402 & 0.4823 & 0.4380 & 0.4717 & 0.4813 \\
    \bottomrule
    \end{tabular}%
  \label{tteststock}%
\end{table}%

\section{Conclusions}
This paper sheds light on the informational efficiency of the corporate bond and stock markets of six EU countries. In particular, we study the evolution over time of the Hurst exponent as a measure of long-range memory using the DFA method. We detect different memory dynamics in stock and bond series. First, the Hurst exponents series are smoother in the stock indices than in the bond indices. Second, the impact of the financial crisis affected more deeply fixed income markets than equity markets. This results could be a consequence of market participants behavior during the financial turmoil. In particular, investors increase their risk aversion, changing their asset allocation. This effect is commonly known as flight to quality and flight to liquidity, i.e. investors prefer safer and more liquid financial instruments. In this case stocks are more liquid than corporate bonds and sovereign bonds are safer than either corporate bonds and stocks. This reaction could generate herd effect that could be reflected in a more persistent time series ($H>0.5$). In spite of the fact that our results cannot be extrapolated to other markets, they have important implications for defining prudential regulation of financial and stock markets. We understand that more research on this topic should be conducted for different markets in order to validate the possible causes of long-range dependence.

\bibliographystyle{plain}      
\bibliography{corporatephysics2}   

\end{document}